\documentclass[sigconf]{acmart} 

\AtBeginDocument{%
  }

\copyrightyear{2025}
\acmYear{2025}
\setcopyright{cc}
\setcctype{by}
\acmConference[IUI '25]{30th International Conference on Intelligent User Interfaces}{March 24--27, 2025}{Cagliari, Italy}
\acmBooktitle{30th International Conference on Intelligent User Interfaces (IUI '25), March 24--27, 2025, Cagliari, Italy}\acmDOI{10.1145/3708359.3712074}
\acmISBN{979-8-4007-1306-4/25/03}

\usepackage{subfigure}
\usepackage{xcolor}
\usepackage{booktabs}
\usepackage{multirow}
\usepackage{setspace}

\newcommand{\rv}[1]{\textcolor{black}{#1}}
\definecolor{myblue}{HTML}{007FFF}
\definecolor{mypink}{HTML}{FF0080}

\begin{document}

\title{Less or More: Towards Glanceable Explanations for LLM Recommendations Using Ultra-Small Devices}

\author{Xinru Wang}
\authornote{Work done during an internship at Meta Reality Labs.}
\email{xinruw@purdue.edu}
\orcid{0000-0002-0213-6425}
\affiliation{%
  \institution{Purdue University}
  \city{West Lafayette}
  \state{Indiana}
  \country{USA}
}

\author{Mengjie Yu}
\email{annaymj@meta.com}
\orcid{0000-0001-6105-1272}
\affiliation{%
  \institution{Meta Reality Labs}
  \city{Redmond}
  \state{Washington}
  \country{USA}
}

\author{Hannah Nguyen}
\email{hannahnguyen@meta.com}
\orcid{0000-0001-9696-4004}
\affiliation{%
  \institution{Meta Reality Labs}
  \city{Redmond}
  \state{Washington}
  \country{USA}
}
\author{Michael	Iuzzolino}
\email{mliuzzolino@meta.com}
\orcid{0009-0006-6445-736X}
\affiliation{%
  \institution{Meta Reality Labs}
  \city{Redmond}
  \state{Washington}
  \country{USA}
}
\author{Tianyi Wang}
\email{tianyiwang@meta.com}
\orcid{0000-0001-9382-6466}
\affiliation{%
  \institution{Meta Reality Labs}
  \city{Redmond}
  \state{Washington}
  \country{USA}
}
\author{Peiqi	Tang}
\email{peiqi@meta.com}
\orcid{0009-0008-2139-0219}
\affiliation{%
  \institution{Meta Reality Labs}
  \city{Redmond}
  \state{Washington}
  \country{USA}
}
\author{Natasha	Lynova}
\email{natashalynova@meta.com}
\orcid{0009-0003-0400-4535}
\affiliation{%
  \institution{Meta Reality Labs}
  \city{Redmond}
  \state{Washington}
  \country{USA}
}
\author{Quoc Co	Tran}
\email{quoct1@meta.com}
\orcid{0009-0000-9068-8934}
\affiliation{%
  \institution{Meta Reality Labs}
  \city{Redmond}
  \state{Washington}
  \country{USA}
}
\author{Ting	Zhang}
\email{tingzhang@meta.com}
\orcid{0000-0001-8156-4809}
\affiliation{%
  \institution{Meta Reality Labs}
  \city{Redmond}
  \state{Washington}
  \country{USA}
}
\author{Naveen	Sendhilnathan}
\email{naveensn@meta.com}
\orcid{0000-0002-3534-890X}
\affiliation{%
  \institution{Meta Reality Labs}
  \city{Redmond}
  \state{Washington}
  \country{USA}
}
\author{Hrvoje	Benko}
\email{benko@meta.com}
\orcid{0000-0002-2059-3558}
\affiliation{%
  \institution{Meta Reality Labs}
  \city{Redmond}
  \state{Washington}
  \country{USA}
}
\author{Haijun	Xia}
\email{haijunxia@ucsd.edu}
\orcid{0000-0002-9425-0881}
\affiliation{%
  \institution{University of California San Diego}
  \city{San Diego}
  \state{California}
  \country{USA}
}
\author{Tanya	Jonker}
\email{tanya.jonker@meta.com}
\orcid{0000-0001-8646-5076}
\affiliation{%
  \institution{Meta Reality Labs}
  \city{Redmond}
  \state{Washington}
  \country{USA}
}
\renewcommand{\shortauthors}{Wang et al.}

\begin{abstract}
Large Language Models (LLMs) have shown remarkable potential in recommending everyday actions as personal AI assistants, while Explainable AI (XAI) techniques are being increasingly utilized to help users understand why a recommendation is given. Personal AI assistants today are often located on ultra-small devices such as smartwatches, which have limited screen space. The verbosity of LLM-generated explanations, however, makes it challenging to deliver glanceable LLM explanations on such ultra-small devices. 
To address this, we explored 1) \textit{spatially structuring} an LLM’s explanation text using defined contextual components during prompting and 2) presenting \textit{temporally adaptive} explanations to users based on confidence levels. We conducted a user study to understand how these approaches impacted user experiences when interacting with LLM recommendations and explanations on ultra-small devices. The results showed that structured explanations reduced users’ time to action and cognitive load when reading an explanation. Always-on structured explanations increased users’ acceptance of AI recommendations. However, users were less satisfied with structured explanations compared to unstructured ones due to their lack of sufficient, readable details.
Additionally, adaptively presenting structured explanations was less effective at improving user perceptions of the AI compared to the always-on structured explanations.
Together with users' interview feedback, the results led to design implications to be mindful of when personalizing the content and timing of LLM explanations that are displayed on ultra-small devices.
\end{abstract}




\maketitle

\section{Introduction}
Artificial Intelligence (AI) advancements over the past decades have led to the development of AI-based personal assistants, which have become a popular way for human users to interact with digital devices and data. 
More recently, Large Language Models (LLMs) such as the GPT \cite{achiam2023gpt4} and LLaMa \cite{touvron2023llama} series have shown remarkable performance in generation tasks, which has led to their growing use as a method to make everyday contextual recommendations for AI-based personal assistants \cite{li2024llm, cosentino2024towards}. 
In addition, to ensure transparency and control during human-AI interaction, there has been a growing interest in Explainable AI (XAI) techniques that can help users understand and trust the AI assistance they receive \cite{ribeiro2016should, zhang2020effect, wang2021explanations, rebanal2021xaigo, xu2023xair}. In the context of LLMs, those models based on transformer architectures are capable of generating explanations that rationalize their reasoning process 
using natural language \cite{wei2022chain, kojima2022large}.

An emerging manifestation of AI-based personal assistant has been on ultra-small devices, such as smartwatches, smartglasses \cite{Meta_2024_orion, google_glass}, and on-cloth gadgets \cite{ai_pin}, to offer low-friction assistance that is seamlessly integrated into the physical world. These devices rely on vast amount of user contextual data gathered from their sensors (e.g., egocentric video, gaze tracking, and audio) to predict the goal of the user's next action and generate action recommendations \cite{jonker2020role}, such as launching an application, to help users efficiently complete daily tasks.

Interacting with LLM recommendations and explanations on such ultra-small devices can, however, be challenging for several reasons.
First, LLM self-explanations are often verbose and time-consuming to comprehend \cite{graphologue}, while the constrained screen space on ultra-small devices exacerbates the challenge of displaying verbose explanations. Real-world examples include smartwatches or Augmented Reality (AR) glasses, which have limited screen real estate that is available, and struggle to display long messages \cite{duchnicky1983readability}. 
Designers need to carefully condense the \textit{content} of an explanation to ensure it remains relevant and informative while not overloading the limited screen space.
Additionally, some tasks, such as providing route adjustment recommendations while driving, are inherently time-sensitive and consume a lot of the user's cognitive resources. Thus, it is crucial to optimize the \textit{timing} of explanations by presenting them only at moments when they would be most useful or when a user requests them to prevent unnecessary user interruptions \cite{xu2023xair}.

To overcome these challenges, this work aims to design glanceable explanations for a goal-oriented, contextual LLM action recommendation system for end-users using ultra-small devices.
We focused on improving \textit{spatial} and \textit{temporal} glanceability via two research questions:

\begin{itemize}
    \item \textbf{RQ1:} To improve the spatial glanceability of LLM explanations on ultra-small devices, how should LLM explanation text be \textit{structured}? 
    
    \item \textbf{RQ2:} To take a step beyond and further improve the temporal glanceability of LLM explanations on ultra-small devices, how should LLM explanations be \textit{adaptively} presented?
\end{itemize}

To answer these questions, we designed a web-based interface that enabled us to simulate an LLM recommending and explaining actions to a user when they were using an AI-based personal assistant embedded in a smartwatch, and we recruited participants to interact with this interface.
A subset of Ego4D \cite{grauman2022ego4d} videos featuring digital interactions in people’s daily activities was selected for evaluation. 
We used a learning approach based on Socratic Models \cite{zeng2022socratic} with the dataset, where pre-trained vision-language models 
were used to generate a linguistic summarization (e.g., user physical actions and detected objects) of a video input for downstream processing with an LLM, 
which then made inferences and generated action recommendations.

To improve spatial glanceability (\textbf{RQ1}), we converted an LLM’s verbose explanation text into concise concepts and intuitive icons that users could grasp at a glance. We adopted a Chain-of-Thought prompt strategy \cite{wei2022chain, kojima2022large} and broke down the inference process into steps. An LLM first summarized all possible contexts (i.e., the \textit{[activity]} the user is doing, the \textit{[object]} the user is interacting with, and the \textit{[location]} the user is in) from the output of the vision-language models. Then, the LLM inferred the short-term \textit{[goal]} that the user wanted to achieve and provided a digital action recommendation that the user could take as a next step. Instead of displaying lengthy text, the explanation only included the four structured components.

To improve temporal glanceability (\textbf{RQ2}), we focused on situations when the intelligent system became uncertain and needed user confirmation \cite{xu2023xair}. 
We employed a hybrid method \cite{xiong2024can, chen2023quantifying} that combined the consistency among multiple responses and the textual confidence generated by the model to extract the confidence level for each recommendation. Explanations were automatically presented whenever the recommendation confidence was low, otherwise they were only displayed upon the user’s request.

We then conducted a user study with 44 participants to understand users’ experiences when provided with  structured and adaptively presented explanations for LLM-driven contextual recommendations. The study was designed with four within-subject conditions: no explanations, always-on unstructured explanations, always-on structured explanations, and adaptive structured explanations whose presentation was dependant on the recommendation confidence level. We found that structured explanations effectively reduced participants’ time to select an AI-recommended action and lowered their cognitive load when reading the explanation. Participants were also more likely to accept the recommendations when presented with always-on structured explanations.
Participants were, however, less satisfied with structured explanations compared to unstructured explanations. When presented adaptively, the structured explanations were less effective at improving user perceptions of the AI (e.g., trust, satisfaction) compared to the always-on structured explanations.
Our analysis of participants’ feedback from a semi-structured interview showed that they valued the ease of reading and viewing text in the structured explanations, but also desired the level of detail and naturalness as provided by the unstructured explanations. Additionally, while participants appreciated the control over interface offered by adaptively presented explanations, it introduced extra efforts to interact with the interface.


Taken together, we make the following contributions in this work: 

\begin{itemize}
    \item A prompting strategy that structures an LLM’s explanation text into defined contextual components to create spatially glanceable LLM explanations on ultra-small devices.
    \item The use of confidence level estimations to adaptively present explanations to create temporally glanceable LLM explanations on ultra-small devices.
    \item A user study to analyze the impact of structuring and adaptively presenting explanations on user experience with AI using ultra-small devices. We conclude by discussing design implications for optimizing the content and timing of glanceable LLM explanations.

\end{itemize}

\section{Related Work}
Our work builds upon three areas of relevant literature, i.e., existing XAI techniques to explain LLMs, empirical studies on AI explanations, and user interface design challenges for ultra-small devices.

\subsection{Techniques to Explain Large Language Models}

A variety of XAI techniques have been developed to increase the interpretability of AI models. Conventional XAI methods can be categorized into model-specific methods tailored to particular model types, and model-agnostic methods applicable across various model types. Training interpretable models such as rule-based models and generalized additive models \cite{jung2017simple, lakkaraju2016interpretable, wang2015falling} are examples of model-specific methods. On the other hand, examples of model-agnostic methods include generating feature contributions \cite{lundberg2017unified, ribeiro2016should, simonyan2013deep} or searching for prototypes or counterfactual instances \cite{kim2016examples, wachter2017counterfactual}.

The emergence of LLMs raised unique challenges when applying traditional XAI methods to explain them due to the size of their training datasets and non-deterministic outputs \cite{liao2023ai}. 
Open-source LLMs like Meta’s LLaMA models \cite{touvron2023llama} provide broader transparency for AI researchers to uncover their internal states. For instance, \cite{voita2023neurons} identified when neurons within LLMs were activated, while \cite{zou2023representation} analyzed how high-level cognitive processes were represented in LLM neurons. \cite{zhu2023physics} explored how LLMs encoded knowledge in their embeddings. 
On the other hand, closed-source LLMs, such as GPT \cite{achiam2023gpt4} and Google Gemini \cite{team2023gemini}, pose more challenges due to the restricted access to their internals. Regardless of the closed or open nature of LLMs, one common approach to generate explanations is to prompt the model to provide “self-rationalizations” alongside predictions \cite{wei2022chain, kojima2022large, lampinen2022can, yao2023beyond}. As these LLMs are based on transformer architectures, they can capture complex relationships between input data and generate output through the attention mechanism, to self-rationalize their recommendations in a way that appears logical to humans. Yet, some work has criticized these self-rationalized explanations as being unreliable and unfaithful to the actual reasoning process \cite{turpin2024language, ye2022unreliability}. To address these limitations, researchers have pursued approaches including iteratively prompting the LLM for self-feedback and refinement \cite{creswell2022selection, madaan2024self}, and grounding an LLM in external knowledge bases \cite{zi2023ierl, chen2023lmexplainer}.

While the natural language modality of LLMs' responses opens up opportunities for explanations that are understandable to non-experts, they are often too verbose to be grasped in a quickly digestible manner. This issue is particularly pronounced on ultra-small devices, and our work aims to address it.

\subsection{User-Centered Design and Evaluation of AI Explanations}

Empirical research has focused on understanding how users understand AI explanations \cite{wang2021explanations, wang2022effects} and trust AI models \cite{zhang2020effect, wang2022effects, wang2023watch}, as well as users' overall task performance \cite{bansal2021does, lai2020why, liu2021understanding, wang2023effects, wang2024human}. This research has also spurred novel design processes that identify users' explainability needs to inform XAI design (e.g., identifying the questions that users commonly ask AI systems \cite{rebanal2021xaigo}). In addition to explaining traditional machine learning models, more recently, user-centered design approaches have been applied to explain large generative AI models. For example, Sun et al. \cite{sun2022investigating} identified explainability needs unique to generative AI for code compared to traditional discriminative machine learning models, and proposed four XAI features: AI documentation, uncertainty indicators, attention visualizations, and social transparency. 

On the other hand, research on designing and evaluating XAI methods for particular digital devices is lacking. \rv{Developing AI explanations for specific interfaces often involves trade-offs between usability and transparency. Springer and Whitaker \cite{springer2019progressive} highlighted that transparency can be distracting and that users might benefit from initially simplified explanations.} Chromik and Butz \cite{10.1007/978-3-030-85616-8_36} surveyed a list of XAI publications and identified design principles for human interaction with explanation user interfaces, including naturalness, responsiveness, flexibility, and sensitivity. Xu et al. \cite{xu2023xair} proposed design principles for displaying XAI information in floating windows on AR devices. \rv{They recommended simplifying explanations by selecting the appropriate content, and triggering explanations only when users have enough capacity, are unfamiliar with the outcome, or when the model is uncertain.}
Our work builds on this prior research but focus on ultra-small devices, where there is an urgent need for highly glanceable explanations.

\subsection{User Interface Design for Ultra-Small Devices}

The ultra-small screen sizes on devices such as smartwatches raises key usability issues for the layout of visually rich content such as verbose text \cite{10.1007/11555261_24, spalink-2002-small, https://doi.org/10.1002/hfm.20733}. To address these challenges, researchers have recommended design guidelines to condense information while maintaining clear user understanding \cite{8b67e137-384d-3dd3-a12e-1ab945cfa108, albers2020information}. For example, a navigation interface should only display an upcoming route as an arrow \cite{10.1145/506443.506514}. Others have explored adaptive menus and have found that users preferred a cloud menu, where predicted items were arranged in a circular tag cloud \cite{10.1145/3237190}. Work by Rahman and Muter showed that presenting continuous text sentence-by-sentence within small display windows offered users the ability to reread one sentence at a time and was found to be as efficient as conventional reading \cite{doi:10.1518/001872099779577264}. In summary, these common approaches included converting lengthy text into graphical diagrams or icons \cite{doi:10.1177/154193120605000508, graphologue}, adaptively prioritizing highly-relevant content \cite{10.1145/1357054.1357249, doi:10.1177/154193120104500602}, progressively displaying content \cite{10.1145/238386.238582, doi:10.1518/001872099779577264}, or utilizing multiple modalities such as voice and gestures \cite{10.1145/2807442.2807500, 10.1145/3010915.3010983, Lee2017}. Guided by these existing design ideas, we selected two major dimensions to explore fitting verbose content on ultra-small screens, i.e., \textit{spatially}, by structuring it into easily viewable formats, and \textit{temporally}, by adaptively displaying the content.

\section{Generating Glanceable LLM Explanations}

In this section, we introduce the dataset and LLM pipeline used for generating recommendations, then outline our approaches to present spatially and temporally glanceable LLM explanations on ultra-small devices. The goal of this pipeline was to imitate a human-AI interface that empowered users to accomplish their daily tasks via a smooth transition to digital actions. The system used observations of longitudinal context as input, which could be any combination of sensors a device had access to, such as egocentric video, biometric data, or digital state information. After predicting the user's goal for their next action based on the longitudinal context, the system recommended a digital action to fulfill this goal and disassembled it into device-executable instructions. Along with the recommendation, an explanation was provided for users to ensure transparency and trustworthiness with the system.

\subsection{Dataset}
We used the Ego4D \cite{grauman2022ego4d} dataset as the testbed to evaluate the LLM pipeline in pre-recorded real-world scenarios. The Ego4D dataset consists of over 3,670 hours of egocentric videos of people’s daily activities. The videos were captured using head-mounted cameras to naturally approximate first-person visual perception. 

We leveraged the annotated video narrations included in the dataset to manually curate a subset of videos that were comprised of interactions with digital devices (e.g., ``phone’’ or ``laptop’’). We trimmed the videos to start 30 seconds prior to the digital action as context for the action and ensured that the video content prior to the digital action contained primarily physical activities, such as cooking or hiking. 
Following this, we were left with 1101 videos 
that focused on everyday goal-oriented, contextual scenarios that the LLM could take as input to generate action recommendations. The daily activities covered all 27 categories of common everyday activities identified by a taxonomist who is one of the authors, such as browsing the internet, chores and cleaning, cooking and eating, exercise, work, gaming, messaging and communication, shopping, and traveling.

\subsection{LLM Recommendation Pipeline}

\begin{figure*}[t]
  \centering
\includegraphics[width=\linewidth]{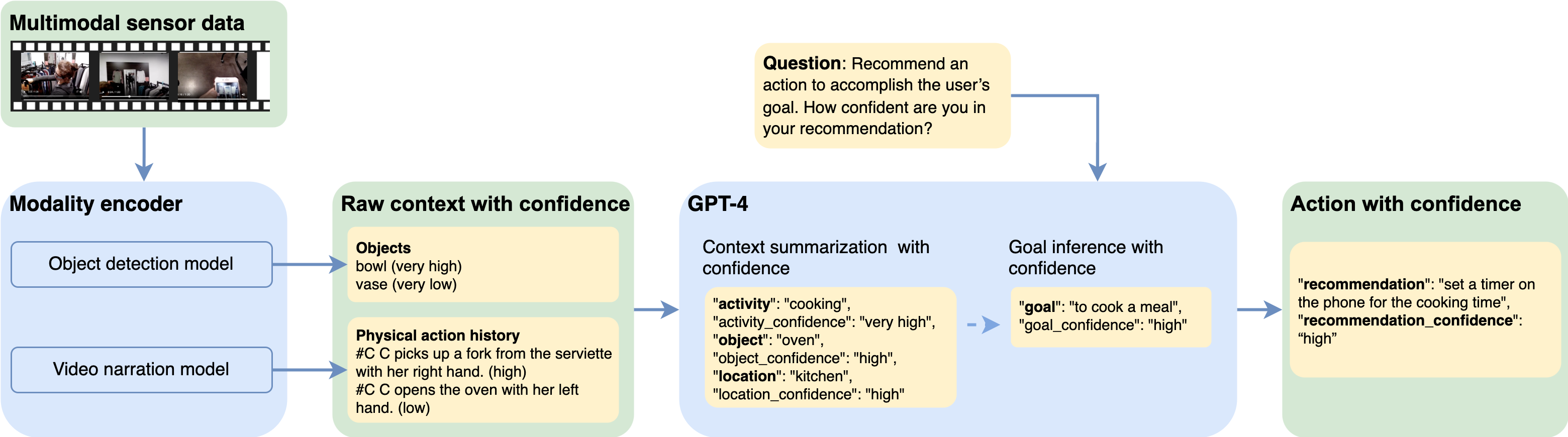}
  \caption{Our LLM pipeline, which was based on Socratic Models, used pre-trained vision-language models 
  to generate a linguistic summary of a video input (detected objects and user physical actions) for downstream processing with an LLM (GPT-4). To obtain the LLM self-explanations, the LLM first summarized all possible contexts (i.e., the [activity] the user is doing, the [object] the user is interacting with, and the [location] the user is in) and then inferred the short-term [goal] that the user may want to achieve. Based on the inferred goal, the LLM then provided a digital action recommendation. Throughout the process, we calculated confidence levels of the output from pre-trained vision-language models. The LLM was then prompted for confidence levels for each contextual component (i.e., [activity], [object], [location], and [goal]) and the recommendation.}
  \label{fig:LLM_pipeline}
\end{figure*}

The pipeline was based on Socratic Models \cite{zeng2022socratic}, where pre-trained vision-language models generated a linguistic summary of a video for downstream processing with an LLM (Figure \ref{fig:LLM_pipeline}). Specifically, we employed 
a pre-trained transformer-based model designed for video understanding 
to extract physical actions (i.e., narrations) present in the video by dividing each video into 2-second clips. We used 
a pre-trained transformer-based object detection model 
to detect objects appearing in the video within a 5-second window before the digital action\footnote{Note that some of the Ego4D videos did not have audio, so only video frames were used as input.}. With these two pieces of contextual information about the video input, GPT-4 was prompted to infer the user's intent and output its recommendation for the user's next digital action.

\begin{figure*}[t]
  \centering
\includegraphics[width=0.95\linewidth]{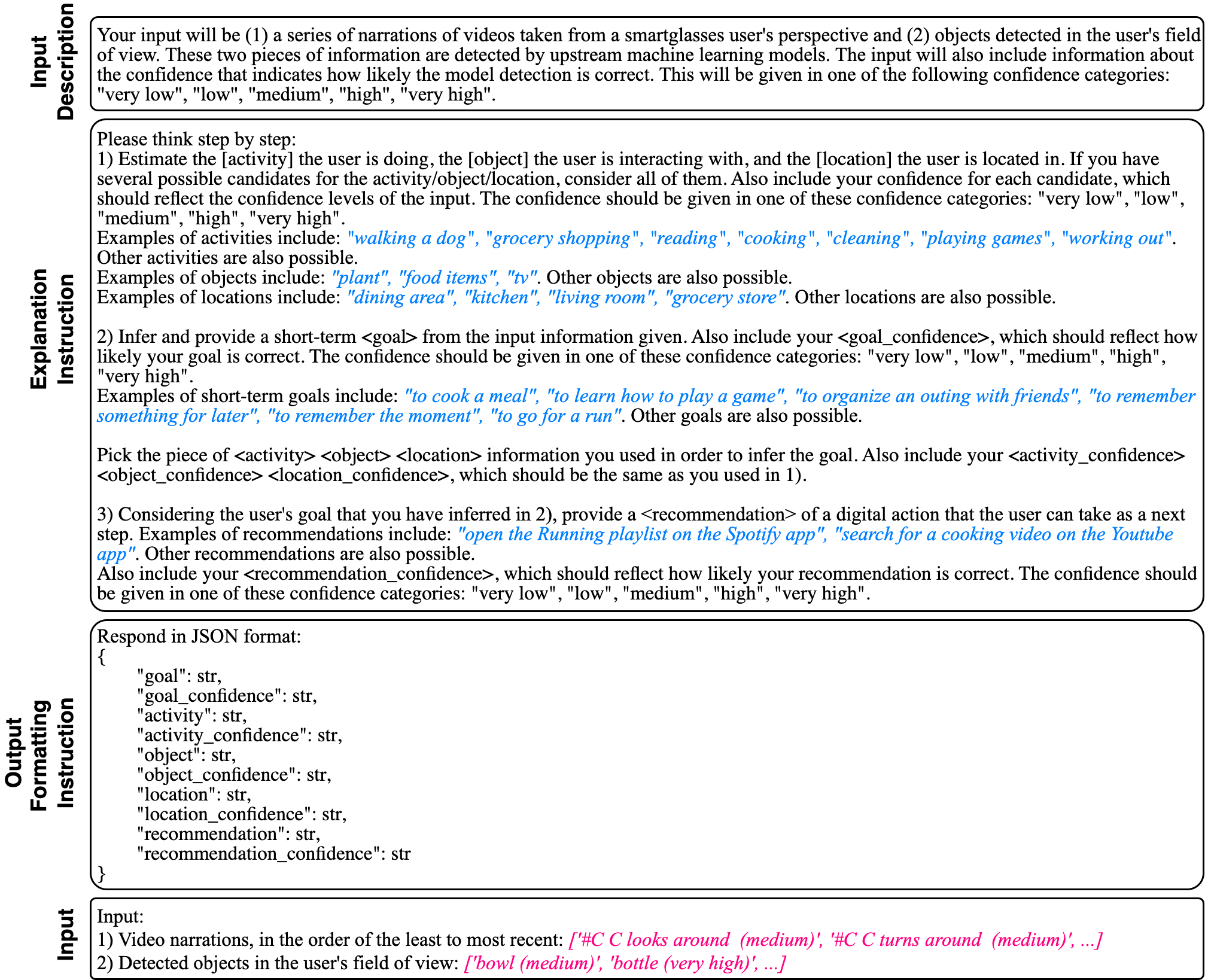}
  \caption{The prompt template included an input description, explanation instructions with few-shot in-context examples (\textcolor{myblue}{\textit{blue text in italics}}), output formatting instructions, and the target input (\textcolor{mypink}{\textit{pink text in italics}}, changed for every query).
  }
  \label{fig:prompt_template}
  \vspace{-10pt}
\end{figure*}

We prompted the LLM to produce natural language explanations along with its recommendations following step-by-step thinking based on the Chain-of-Thought method \cite{wei2022chain, kojima2022large}. 
As text-based responses from LLMs can be too verbose to be displayed on an ultra-small device's screen, spatially, we structured the textual responses using defined contextual components to create a visual experience that required minimal cognitive effort to comprehend. Temporally, we controlled the presentation timing of the explanations to avoid unnecessary interruptions to the user. We addressed situations where the AI system showed uncertainty, meaning it might make errors and require user confirmation. The prompt template is shown in Figure \ref{fig:prompt_template}.

\subsubsection{Structuring Explanation Text Using Defined Contextual Components}

\label{sec:q1_what}
The inference process was divided into several stages. The LLM first extracted and summarized the relevant contextual information, i.e., the [\textit{activity}] the user was performing, the [\textit{object}] the user was interacting with, and the [\textit{location}] of the user, from the output generated by pre-trained vision-language models. Then, the LLM inferred the user's short-term [\textit{goal}] based on these contextual cues. Finally, using the inferred goal, the LLM provided a digital action recommendation.
We chose ``object'', ``activity'', ``location'', and ``goal'' as the contextual information as these entities were the core contextual components necessary to generate relevant recommendations and explanations in an everyday mixed-reality scenario, as provided by the aforementioned taxonomist.
We also referenced definitions and few-shot examples in the prompt for each component. The LLM was instructed to provide its output in a JSON format, which was later used to create the structured representations.

\subsubsection{Adaptively Presenting 
Explanations Based on Confidence Levels}

Following Chen and Muelle \cite{chen2023quantifying} and Xiong et al. \cite{xiong2024can}, we employed a hybrid approach that combined consistency-based methods and verbalized confidence to extract a calibrated confidence level for each recommendation. We mapped the confidence score for each detected object from the object detection
model and the perplexity score for each detected physical action from the 
video narration model into one of five textual confidence levels (i.e., ``very low'', ``low'', ``medium'', ``high'', ``very high''), based on the quantile position of the score. We opted for a textual representation of confidence to ensure a unified representation of “confidence”, as well as to maximize the LLMs’ abilities to comprehend natural language. We included the confidence levels for raw contextual information in the prompt, then prompted the LLM to output confidence levels along with the recommendation. 

We then generated a reference response $y_0$ and confidence $c_0$ with temperature sampling set at 0. Using the same prompt, we produced $K=5$ candidate responses $\{y_1, y_2, ..., y_K\}$ and their verbalized confidences $\{c_1, c_2, ..., c_K\}$ by increasing the temperature value to 0.7 \cite{xiong2024can}. Each candidate confidence $c_i$ ($i \in \{1,2,...,K\}$) was transformed back into a numerical value and was then updated by incorporating the reference answer and the similarity to the reference answer, denoted as $s_i$: $\bar {c_i} = \frac{(c_0+c_i)}{2}s_i$ ($i \in \{1,2,...,K\}$). The similarity scores were obtained using the BERTScore model \cite{Zhang2020BERTScore:}, which measured the semantic similarity between two texts. Finally, we computed the average contribution from all candidate answers as the final confidence score $c_{hybrid} = \sum_{i=1}^{i=K}\bar {c_i}/K$, and then converted it into one of the five confidence levels based on the quantile position of the score.

\subsection{Technical Evaluation}
\label{sec:tech_eval}
To validate the quality of structured explanation components and confidence levels that the system generated, we conducted a small-scale technical evaluation. We randomly sampled one video from each of the 27 activity context categories. Three coders then coded how likely it was that the AI recommendation was correct, how likely it was that the inference about each explanation components (i.e., activity, object, location, and goal) was correct (i.e., plausibility), and to what extent each explanation component supported the AI recommendation (i.e., faithfulness) \cite{mathew2021hatexplain, el2022evaluation}, for each of the 27 videos using a 7-point Likert scale. 

We then calculated the inter-rater agreement across the coders using Krippendorff's $\alpha$ and obtained a score of 0.553, which is considered acceptable for exploratory research involving subjective ratings \cite{krippendorff2018content}. We found that all explanation components were rated as plausible and faithful to the recommendation, with the median of all ratings being larger than 4 (one-sample Wilcoxon signed-rank test showed $p<0.05$ for all explanation components for the plausibility and faithfulness measures; Table \ref{tab:tech_explanation}). 
To assess the calibration of the recommendation confidence, we computed the Pearson’s $r$ correlation coefficient between the coders' estimations of the recommendation correctness likelihood (as ground truth) and the hybrid confidence $c_{hybrid}$, obtaining a significant positive coefficient of $r = 0.559$ ($p<0.01$). 
Meanwhile, the verbalized confidence levels $c_0$ did not correlate with the coders' estimations ($r=0.171$, $p=0.393$), indicating that the hybrid approach could lead to more calibrated confidence.

\begin{table}[t]
\caption{Ratings for the plausibility and faithfulness of each explanation component. 
}
\label{tab:tech_explanation}
\begin{center}
\begin{tabular}{lcc}
\toprule
\begin{tabular}[c]{@{}l@{}}\textbf{Explanation}\\\textbf{Component}\end{tabular} & \begin{tabular}[c]{@{}l@{}}\textbf{Plausibility}\\median (IQR)\end{tabular} & \begin{tabular}[c]{@{}l@{}}\textbf{Faithfulness}\\median (IQR)\end{tabular} \\
\midrule
Goal  & 5.67 (3.50 - 6.67)   & 6.67 (6.50 - 7.00)\\
Activity  & 6.67 (6.67 - 7.00)  & 6.333 (5.50 - 6.67)\\
Object   & 7.00 (6.67 - 7.00)  & 6.333 (5.00 - 7.00)\\
Location  & 6.67 (3.00 - 7.00) & 5.000 (3.33 - 6.67) \\
\bottomrule
\end{tabular}
\end{center}
\vspace{-10pt}
\end{table}

\section{User Study Design}
To gain insights into how LLM explanations that differ in their structure and adaptivity influence user experiences, we conducted a lab-based user study. The study measured participants' perceptions of LLM recommendations and explanations while they watched videos of everyday interactions and received recommendations on a simulated smartwatch UI.

\subsection{Participants}
Forty-four participants (Male = 20, Female = 23, prefer not to say = 1; mean age = 38 years, std = 12.3 years) were recruited to participate in our in-person user study, a sample size comparable to prior work \cite{caine2016local}. 
Study participants were recruited through email invitations and social media platforms from an existing participant pool. Twenty participants had a graduate degree, 18 had a Bachelor's degree, 4 had some college experience, and 2 did not have any education past high school. Nine participants worked on AI-powered devices or products, 11 used AI-powered devices or services at least once a week, 9 used them at least once a month, 9 had used them a few times, and 6 had never interacted with AI-based devices or services. The entire study lasted about one hour, and each participant was compensated \$75 upon completion of the study.

\subsection{Task}
\begin{figure*}[t]
  \centering
\includegraphics[width=0.65\linewidth]{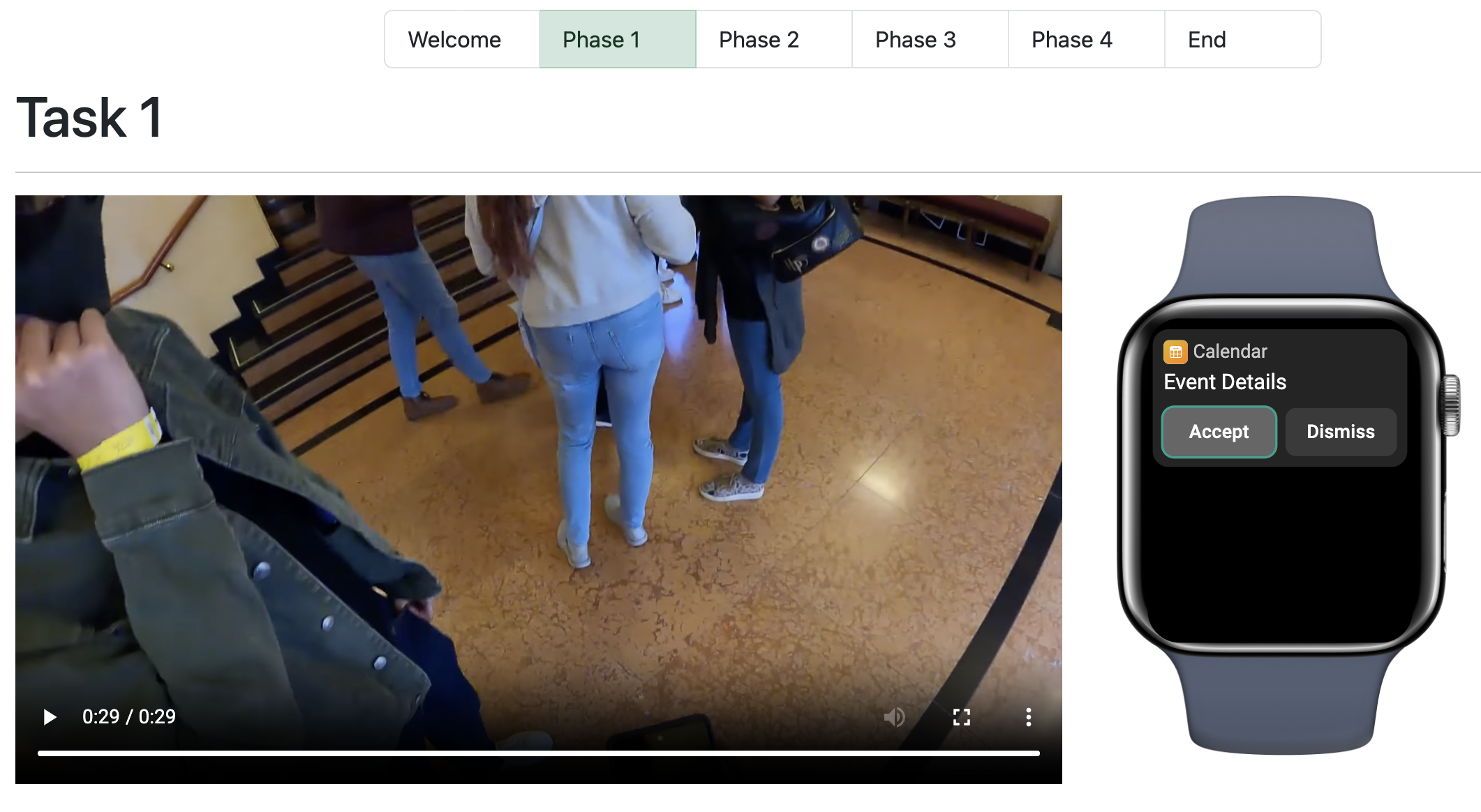}
  \caption{The interface that participants saw on the desktop computer. On the left, they could watch the 30-second Ego4D video. They could chose whether to accept or dismiss the AI’s recommendation on the smartwatch UI on the right.}
  \label{study_interface}
\end{figure*}

During the study, participants were asked to watch a 30-second Ego4D video on a standard desktop computer and were told that the videos they were watching were captured by a camera embedded in a pair of smart glasses that were worn on a person’s head and naturally approximated the visual field of the wearer of the glasses. We instructed participants to imagine that they were the one wearing these smartglasses and they were performing the activities shown in the video. They were also told that while they were imagining performing these activities, their personal AI assistant would try to predict their next intent and recommend a digital action, which was pushed to their smartwatch. The smartwatch UI was displayed next to the video on the desktop computer. Participants were asked to evaluate whether the recommendation met their needs (i.e., as the wearer of the glasses in the video) and then choose to accept or dismiss the AI's recommendation on the smartwatch UI by clicking on it (Figure \ref{study_interface}).

We opted to build a web-based interface that simulated smartwatch UI for two reasons. 
First, replicating a wide variety of real-life scenarios in a lab environment would be difficult. Instead, the existing Ego4D dataset provided a diverse set of pre-recorded situations.
Second, our recommendation pipeline, which employed large transformer-based models (i.e., 
vision-language models, GPT-4), had slow response times. Deploying it on a smartwatch for real-time video processing would cause significant latency (\textasciitilde tens of seconds delay), which would have influenced participant satisfaction.

\subsection{Conditions}
\begin{figure}[t]
  \centering
  \subfigure[No explanation]{
\includegraphics[width=0.28\linewidth]{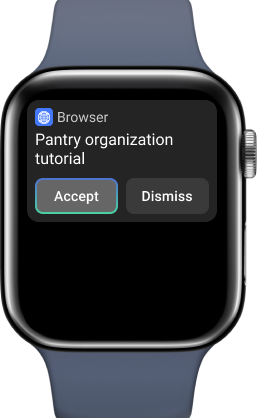}
\label{fig:T0}
    }
    \hspace{6pt}
    \subfigure[Always-on unstructured explanation]{\includegraphics[width=0.28\linewidth]{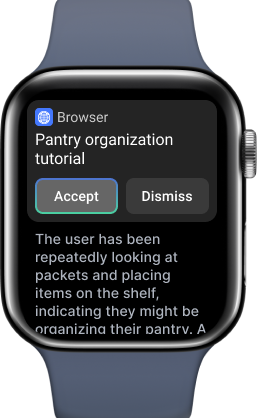}
    \label{fig:T1}
    }
    \hspace{6pt}
    \subfigure[Always-on structured explanation]{
\includegraphics[width=0.28\linewidth]{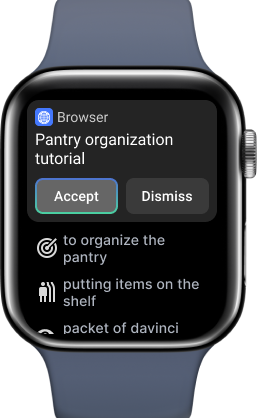}
\label{fig:T2}
    }\\
    \subfigure[Adaptive structured explanation (low confidence)]{
    \includegraphics[width=0.28\linewidth]{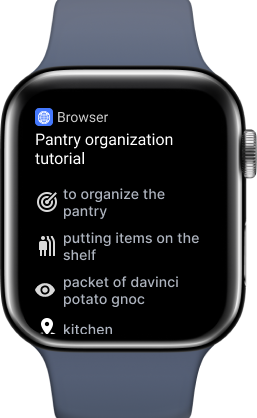}
\label{fig:T3}
    }
    \hspace{6pt}
    \subfigure[Adaptive structured explanation (high confidence - default)]{
    \includegraphics[width=0.28\linewidth]{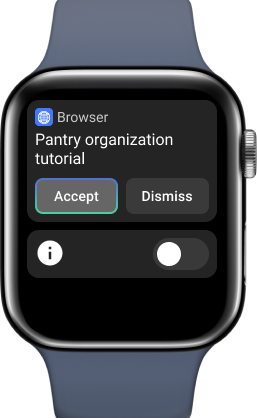}
\label{fig:T3_high0}
    }
    \hspace{6pt}
    \subfigure[Adaptive structured explanation (high confidence - toggled)]{
    \includegraphics[width=0.28\linewidth]{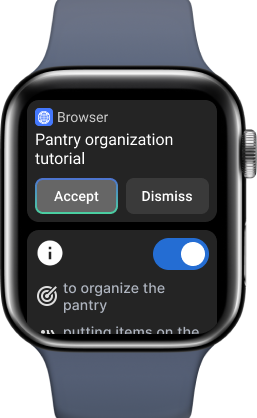}
\label{fig:T3_high1}
    }
  \caption{The Smartwatch UI designs for the four conditions in the user study. (d)-(f) illustrate different variations of the adaptive structured explanation condition.}
  \label{fig:watch}
\end{figure}

A within-subject design was used with four conditions that manipulated whether and how the explanations were displayed (Figure \ref{fig:watch}). A balanced 4 $\times$ 4 Latin Squares design was used for condition counterbalancing \cite{mackenzie2013human}. A UX designer who is one of the authors standardized the smartwatch UI design 
using Figma.

\begin{itemize}
    \item In the \textbf{no explanation} condition (Figure \ref{fig:T0}), participants received recommendations without any explanation on the smartwatch UI. 
    \item In the \textbf{always-on unstructured explanation} condition (Figure \ref{fig:T1}), participants received recommendations along with an unstructured explanation on the smartwatch UI. The number of words in the unstructured explanation were limited to ensure they occupied the same amount of screen space as the structured explanations (i.e., the following two conditions), while allowing participants to scroll up and down to view more text. The prompt used to generate the unstructured explanations can be found in Appendix \ref{app:prompt_baseline}. 
    \item In the \textbf{always-on structured explanation} condition (Figure \ref{fig:T2}), participants received a recommendation along with a structured explanation that was comprised of four components: a goal, an activity, an object, and a location. 
    \item In the \textbf{adaptive structured explanation} condition, structured explanations were provided adaptively depending on the recommendation's confidence level. The explanation was displayed automatically when the recommendation confidence was low (Figure \ref{fig:T3}), otherwise, it was hidden by default (Figure \ref{fig:T3_high0}) and a toggle button enabled participants to see it ((Figure \ref{fig:T3_high1})). 
\end{itemize}

For the adaptive structured explanation condition, we categorized recommendation confidence levels into a binary classification as high or low, depending on whether the $c_{hybrid}$ level was “low”/“very low” or “high”/“very high". These categorizations were based on the findings from our technical evaluation (Section \ref{sec:tech_eval}), where we observed a clear gap in the human estimation of the recommendation correctness likelihood based on the confidence levels ($c_{hybrid}$). When $c_{hybrid}$ was ``low'' (median correctness rating = 3.5) or ``very low'' (median correctness rating = 3.0), coders rated the correctness likelihood much lower compared to recommendations with $c_{hybrid}$ being ``high'' (median rating = 6.0) or ``very high'' confidence levels (median rating = 5.5). \rv{From the same pool of task instances used in the technical evaluation described in Section \ref{sec:tech_eval}}, we chose 20 task instances where the recommendation confidence $c_{hybrid}$ was “low”/“very low” and 20 instances where the confidence $c_{hybrid}$ was “high”/“very high.” To ensure the representativeness of the selected instances, we sampled all task instances based on the data distributions across the 27 activity context categories in the video dataset.

\subsection{Study Procedure}
Upon arrival at our lab, participants were asked to complete a consent form and an initial survey about their demographics and experience with AI. 

Participants then completed 40 trials, which were divided into four blocks, one block for each condition. Before beginning each block, there were two tutorials that were identical to the actual trials but during which no user data was collected. Each block then consisted of 10 trials, with half of them having low confidence recommendations and half having high confidence recommendations. The order of all 10 trials within each block was randomized across participants.  During each trial, participants watched a 30-second video on the desktop computer. At the end of the video, a recommendation for a digital action and an explanation (depending on the assigned condition) was displayed on the simulated smartwatch UI. Participants then chose whether to accept or dismiss the AI’s recommendation. After each block, participants completed a survey about their experience with the AI's recommendations and explanations.

Upon completion of all four blocks, participants completed a final survey and answered interview questions to understand what they liked or disliked about each AI explanation condition, when they would like to see an explanation, what information they would like to see in the explanation, and how the AI explanation could be improved to better meet their needs. 

\subsection{Metrics}

Several metrics were computed to understand participants’ experiences and perceptions when presented with the AI explanation conditions. First, we computed the following objective measures:
\begin{itemize}
    \item \textbf{Time to action}: The time taken by a participant from the end of the video until they accepted or dismissed the AI recommendation.
    \item \textbf{Acceptance rate}: The percentage of tasks where the participant chose to accept the AI recommendation
\end{itemize}

At the end of each block, we asked participants about the following subjective measures using 7-point Likert scales:

\begin{itemize}
    \item \textbf{Mental load}: We measured participants' mental load while making the decision, understanding the recommendation, and reading the explanation via three questions: 
    \begin{itemize}
        \item ``\textit{Overall, how much mental effort did you spend on deciding whether to accept or dismiss the AI recommendation?}'' 
        \item ``\textit{Overall, how much mental effort did you spend on understanding how the AI model makes recommendations based on the context in the video?}'' 
        \item ``\textit{How much mental effort did you spend on reading the AI model’s explanation of the recommendation?}''
    \end{itemize}
    \item \textbf{Trust in AI}: Participants' trust in AI was calculated by the average rating of the following three questions: \begin{itemize}
        \item ``\textit{I have faith that the AI model would be able to cope with all different situations.}'' 
        \item ``\textit{I am confident that the AI model can make good recommendations.}'' 
        \item ``\textit{Recommendations made by the AI model are likely to be reliable.}''
    \end{itemize}
    \item \textbf{Understanding of AI.}: This was also computed by averaging participants' responses to three questions:
    \begin{itemize}
        \item ``\textit{I understand how the AI model works to predict my next digital action based on the context in the video.}'' 
        \item ``\textit{I can predict how the AI model will behave based on the context in the video.}'' 
        \item ``\textit{I feel the AI model is transparent in communicating how the recommendations were made based on the context in the video.}''
    \end{itemize}
    \item \textbf{Satisfaction with AI}: ``\textit{I’m satisfied with the AI model’s recommendation.}''
    \item \textbf{Satisfaction with AI explanations.}: This metric differed from ``satisfaction with AI'' as it specifically measured how participants felt about the \textit{explanation} provided by the AI for its recommendations, i.e.,  ``\textit{I’m satisfied with the AI model’s explanation of  the recommendation.}''
\end{itemize}

At the end of all four blocks, participants also ranked the four conditions based on their \textbf{overall preferences} for them. 


\subsection{Analysis Methods}

Mixed-effect regression models were used to analyze the aforementioned metrics. For the time to action and acceptance rate, each participant and trial were random effects, while the conditions were fixed effects. For the mental load, trust, satisfaction with the AI, understanding of the AI, and the satisfaction of the AI explanations, each participant was a random effect. For the analysis of participants' rankings of the four conditions, a cumulative link mixed-effects model was used to analyze the ordinal response variable. The results of these models were interpreted through the estimated coefficient values for the fixed effect variables. 

The interview data was analyzed by two coders using thematic analysis \cite{braun2006using}. Two authors iteratively discussed and developed a codebook. The inter-rater reliability was calculated using Cohen's kappa \cite{mchugh2012interrater}, where $\kappa=0.82$ ($\text{SE}=0.055, \text{95\% CI}=[0.71, 0.93]$). The final themes were developed and refined over multiple iterations of discussion.

\section{User Study Results}
We analyzed participants' time to action, acceptance rate, subjective perceptions, and interview data to better understand their experiences with the different AI explanation conditions. 

\subsection{Quantitative Results}

\subsubsection{Time to Action}

\begin{figure}[t]
  \centering
    \includegraphics[width=0.8\linewidth]{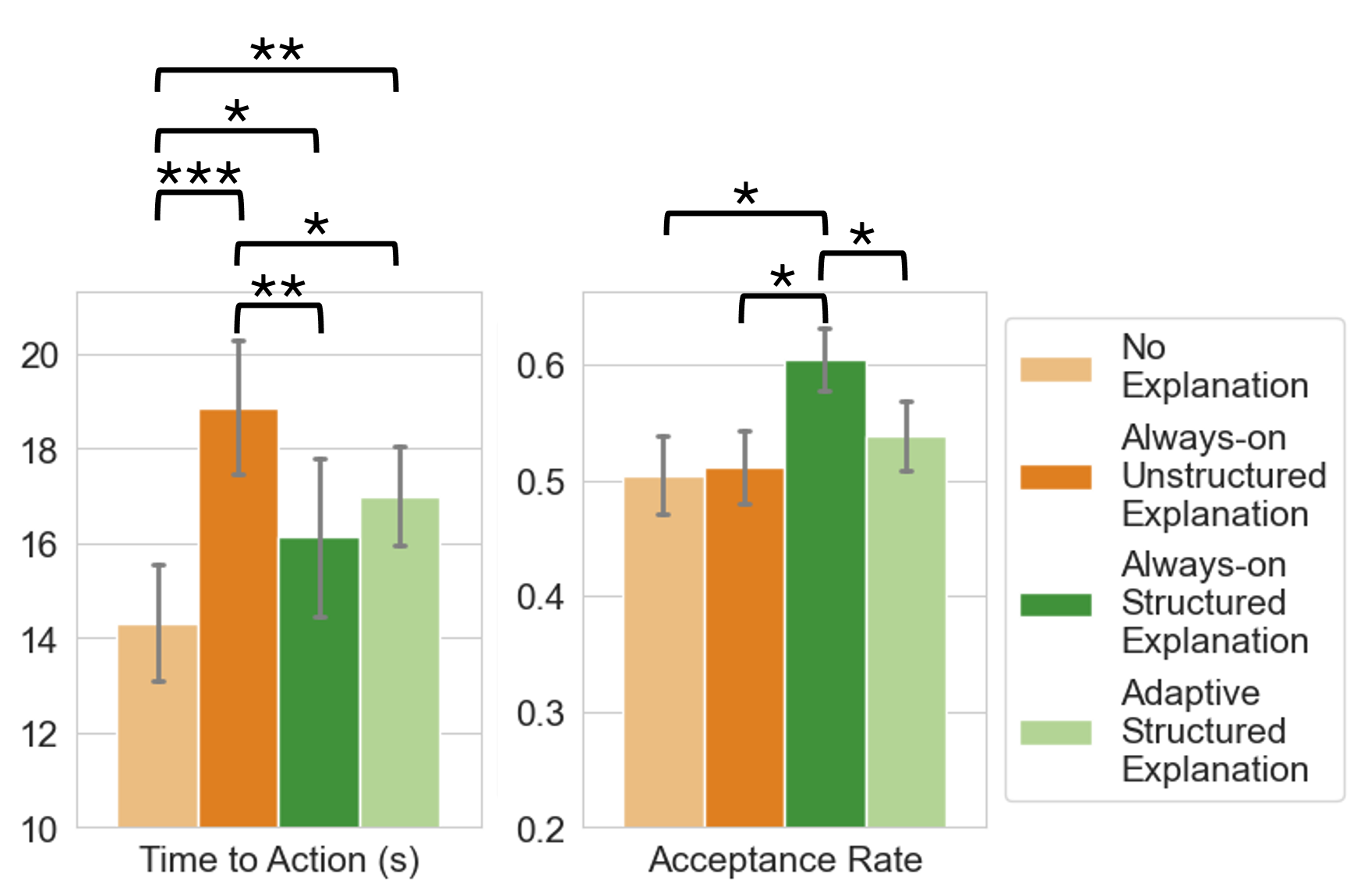}
  \caption{Participants’ time to action and acceptance rate for each AI explanation condition. The error bars represent standard errors. (*: $p$ < 0.05; **: $p$  < 0.01; ***: $p$  < 0.001)}
  \label{fig:time+acceptance}
\end{figure}

Our results showed the presentation of the explanations impacted user's time to action. Compared to the no explanation condition, providing explanations significantly increased the time participants took to select an AI recommendation (always-on unstructured: $\beta=4.6467, p<0.001$, always-on structured: $\beta=1.9341, p<0.05$, always-on unstructured: $\beta=2.7492, p<0.01$). But among the three conditions with explanation, structured explanations---both always-on structured explanations ($\beta=-2.7126, p<0.01$) and adaptive structured explanations ($\beta=-1.8975, p<0.05$)---significantly reduced participants' time to action compared to unstructured explanations (Figure \ref{fig:time+acceptance}).


\subsubsection{Acceptance Rate}
Our results found that always-on structured explanations resulted in a higher acceptance rate compared to the other three conditions (compared with no explanation: $\beta=0.0675, p<0.05$, with unstructured explanations: $\beta=0.0712, p<0.05$, and with adaptive structured explanations: $\beta=0.0584, p<0.05$; Figure \ref{fig:time+acceptance}). 


 \subsubsection{Mental Load}
\begin{figure*}[t]
  \centering
\includegraphics[width=\linewidth]{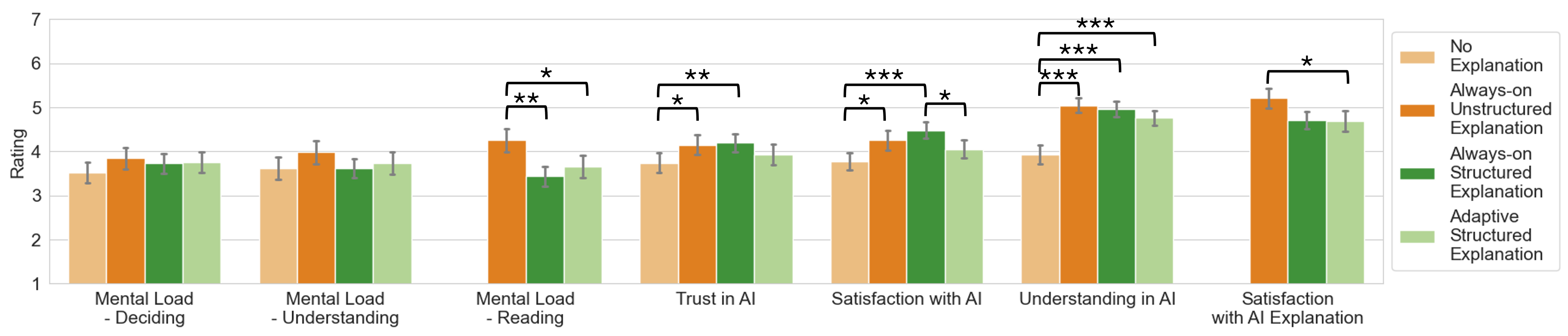}
    
  \caption{Participants’ reported mental load, trust, satisfaction, understanding, and satisfaction with the AI Explanations. The error bars represent standard errors. (*: $p$ < 0.05; **: $p$  < 0.01; ***: $p$  < 0.001)}
  \label{fig:result_sub}
\end{figure*}

 We found no significant differences in participants' mental load when deciding or understanding recommendations (Figure \ref{fig:result_sub}), however, participants rated their mental load as significantly lower when reading structured explanations (i.e., always-on structured: $\beta=-0.8182, p<0.01$, adaptive structured: $\beta=-0.5909, p<0.05$) than when reading unstructured explanations, which aligns with the time to action findings.

\subsubsection{Trust in AI}
We found that both the always-on unstructured explanations ($\beta=0.4091, p<0.05$) and always-on structured explanations ($\beta=0.4545, p<0.01$) led to higher levels of trust when compared to the no explanation condition (Figure \ref{fig:result_sub}). There was no difference between the always-on unstructured, always-on structured, and adaptive structured explanation conditions.

\subsubsection{Satisfaction with AI}
When presented with always-on unstructured and always-on structured explanations, participants were both more satisfied with the recommendations given by the AI compared to when no explanation was provided (always-on unstructured: $\beta=0.4773, p<0.05$; always-on structured: $\beta=0.7045, p<0.001$; Figure \ref{fig:result_sub}). The use of adaptive structured explanations did not result in participants being more satisfied with the AI's recommendations compared to when no explanation was provided, and even led to significantly lower satisfaction compared to the always-on structured explanation condition ($\beta=-0.4318, p<0.05$).

\subsubsection{Understanding of AI}
When participants were provided with an explanation, they had a better understanding of the rationale behind the AI recommendation than when they were not (e.g., always-on unstructured explanations: $\beta=1.1136, p<0.001$, always-on structured explanations: $\beta=1.0303, p<0.001$, always-on structured explanations: $\beta=0.8258, p<0.001$; Figure \ref{fig:result_sub}). There was no significant difference between the always-on unstructured, always-on structured, and adaptive structured explanation conditions, suggesting that structuring and adaptively presenting the unstructured explanation did not negatively impact participants' understanding of the AI.

\subsubsection{Satisfaction with AI explanations} 
We also found that adaptive structured explanations were less satisfying compared to always-on unstructured explanations ($\beta=-0.5227, p<=0.05$; Figure \ref{fig:result_sub}). Though not statistically significant, always-on structured explanations were rated as marginally less satisfying than always-on unstructured explanations ($\beta=-0.5000, p=0.053$). 

\subsubsection{Overall Preferences}
\begin{figure}[t]
  \centering
\includegraphics[width=\linewidth]{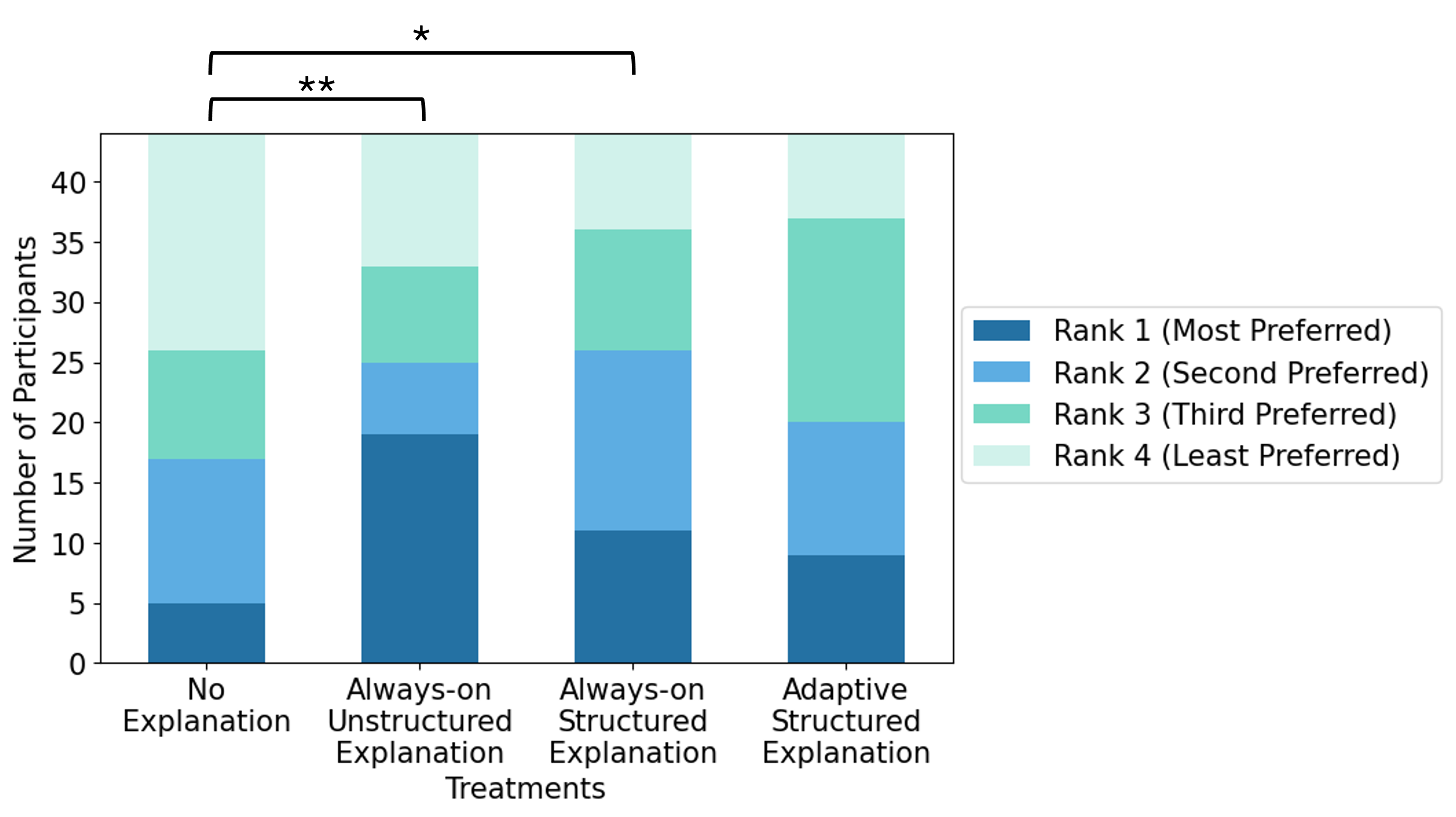}
\vspace{-10pt}
  \caption{Participants’ preference rankings for the four AI explanation conditions.}
  \label{fig:overall_pref}
\end{figure}

In terms of participants' overall preferences for the four conditions (Figure \ref{fig:overall_pref}), both the always-on unstructured explanations ($\beta=-1.1543, p<0.01$) and the adaptive structured explanations ($\beta=-0.9242, p<0.05$) were ranked higher than the no-explanation condition. 
However, the difference in ranks between the adaptive structured explanation condition and the no explanation condition was not statistically significant ($\beta=-0.6752, p=0.074$).
Additionally, no significant difference was found between the three explanation conditions. 

\subsection{Qualitative Results}

Seven themes were identified within the interview feedback. Some of the themes helped explain why participants preferred some conditions over others, whereas others speculated about other ways that explanations could be improved. Each theme is elaborated in the following sections.

\subsubsection{Usability} 
Usability referred to the amount of effort or time to interact with interface, i.e., whether the interface was easy and quick to use, or required more effort. Participants generally wanted to perform minimal actions and reading on the interface, with easy viewing and decision-making. For example, some participants noted that they would only want explanations when they had enough cognitive capacity, e.g., ``\textit{If it were detecting locations or the amount of activity I was doing, then I'd probably prefer the simpler interface. If I were driving or doing something active, it wouldn’t be helpful for me to stop and read that explanation, so I’d dismiss it.}'' (P13) 

Twenty participants valued the high usability of the no explanation condition, saying the interface was easy and fast to use, e.g., ``\textit{It allowed me to react fastest}'' (P10). In conditions with explanations, eight participants mentioned that the information provided by always-on structured explanations was easy to access, e.g., ``\textit{The explanations were automatically expanded}'' (P13). The adaptive structured explanations received mixed feedback. Fifteen participants found them easy and fast to use (e.g.,``\textit{[They] were very simple and to the point}'' (P2)). On the other hand, usability concerns were raised by 30 participants who disliked the extra effort needed to interact with the interface (e.g.,``\textit{You have to hit an extra button, which just seems like an unnecessary step}'' (P29)).

\subsubsection{Sufficiency}
Participants also noted that sufficiency, or the amount of information required to understand AI-generated recommendations or explanations, was important. This included both wanting more information and removing unnecessary details. Thirty participants mentioned their desire for more transparency in the AI decision process in the no explanation condition to help them understand AI decision better (e.g., ``\textit{it doesn't have any explanation}'' (P37)). Twenty-four participants favored the comprehensive details provided by the unstructured explanations, e.g., ``\textit{it gives you a more comprehensive explanation as to what's happened}'' (P9). Regarding the structured explanations, 18 participants thought the four icons contained sufficient information, whereas 15 participants suggested that the icons could have been less redundant and location information was unnecessary. 
For example, ``\textit{the feedback I got just like seemed redundant. It just kind of felt like a waste of time to read down a list of four points that often had the same information in them}'' (P33).

\subsubsection{Readability}
Readability referred to the ease of reading and viewing text or icons presented on the interface (e.g., ease of viewing information, too much text, low cognitive load). The comparison primarily focused on unstructured versus structured explanations. Thirty-one participants noted that the unstructured, free-text explanations resembled human speech and were easy to understand (e.g., ``\textit{it would be if like a human was just explaining it to me, like, oh because I saw that this and so I predicted your in this environment trying to do this ... I feel that's easy to understand}'' (P40). The downside, as complained by 30 participants, was that there was too much text making the explanations difficult to read (e.g., ``\textit{It was hard to read this block of text}'' (P7).

On the other hand, 8 participants mentioned that the structured explanations were easy to view (e.g., ``\textit{it was just really nicely laid out, organized}'' (P22). Twenty-three participants, however, found the icons and the logic between them difficult to understand (e.g., ``\textit{I had to like keep reminding myself what they meant}'' (P1) and ``\textit{those four icons didn't always connect}'' (P12)).

\subsubsection{User Control}

Participants also highlighted the importance of the amount of control user they had over the interface and their choices (e.g., giving users more options, allowing the user to decide when to view information). They believed that the AI ``\textit{should give you that option ... ask you, would you like more suggestions? }'' (P21). Twenty-eight participants expressed a desire for more control over explanation length, appreciating the greater user control offered in the adaptive structured explanation condition to toggle explanations on or off (e.g., ``\textit{[this condition] was good in a way that they have an option, have it on or have it off}'' (P9)).

\subsubsection{Accuracy}
Accuracy concerns related to the contextual relevance of AI-generated recommendations and information. 
Twenty-four participants wanted more accurate recommendations for certain situations and sometimes found that recommendations or explanations didn’t match the context (e.g., ``\textit{The context of the video was totally different from the action that I'm kind of like being asked to perform}'' (P30)).

\subsubsection{Multi-Modality}

 Participants thought that explanations should support options for other modes of accessibility, such as information presented in multiple languages, audio input/output, or visual display. This was mentioned by 19 participants as a potential way to improve the experience. For instance, one of the videos shown in the study contained dialogue spoken in Russian, which provided direct information on the next recommended action, but the AI assistant failed to capture it. 
Another example came from P37, who explained, ``\textit{if I don't need to read but just show me some pictures or images like for example, for objects you don't need to mention like what object it is, you just put what I'm looking at the picture here.}''
 
\subsubsection{Personalization}

Lastly, 43 participants showed interest in personalization. They thought that explanations should contain information customized for personal preferences, such as providing recommendations based on daily routine and previous actions, displaying explanations for unfamiliar tasks, and being less frequent for tasks users are already familiar with. As P12 suggested, ``\textit{one thing I think would be helpful to add [to improve the explanation] could be like historical action.}''

\subsection{Summary}

In this section, we bridge the qualitative and quantitative results by showing how the qualitative feedback from participants provides context for interpreting the quantitative results.

\textbf{Structured explanations are quick and easy to use, but fall short in providing sufficient details and the naturalness of human-like language.}
The quantitative results showed that both always-on structured explanations and adaptive structured explanations reduced participants' time to action and mental load when reading the explanations, compared to unstructured explanations. Interview feedback supported these findings, as participants mentioned that structured explanations were clear to view and access (i.e., high usability), allowing them to quickly glance at the information, which contributed to shorter time to action. This may also explain why always-on structured explanations also resulted in a higher acceptance rate of AI recommendations compared to all other conditions, i.e., they had less time to overthink and instead relied on the easily-accessible icons, which reinforced their belief that the recommendation seemed trustworthy.

Despite these advantages, participants rated the always-on structured explanations (marginally) and adaptive structured explanations as less satisfying than the always-on unstructured explanation. During the interviews, participants mentioned that structured explanations were clear to view, but they struggled to remember the meaning of the icons and their logical connection. Unstructured explanations, on the other hand, offered more comprehensive and human-like information. 
\rv{The process of mapping abstract icons to their semantic meanings likely imposed an extraneous cognitive burden, which may have led participants to value ``sufficiency'' and ``readability'' more when evaluating whether an explanation was satisfactory, thus leading to higher satisfaction ratings for unstructured explanations.}

\textbf{Adaptive structured explanations offer users control over the interface, but the (inappropriate) extra toggling hindered effective human-AI interaction.}
In comparison to the no explanation condition, all three explanation conditions led to a higher understanding of the AI. However, the adaptive structured explanation condition---where participants needed to proactively interact with the interface to view details---was less effective in showing the benefits of providing AI explanations. Specifically, both always-on unstructured and always-on structured explanations increased participants' trust and satisfaction in the AI's recommendations and were preferred over the no explanation condition, but this was not observed for the adaptive structured explanations. The qualitative feedback revealed that while participants appreciated having control over when to view explanations, the additional toggle action felt like a redundant barrier that impaired the smoothness of  their interaction with the interface, thus \rv{shifting their focus from quickly grasping the information to managing the interface.}

\section{Discussion}
Findings from our user study drive the design implications for creating glanceable AI explanations on ultra-small devices that meet users' expectations and needs more effectively.  In this section, we discuss these design implications as well as the limitations of our work.

\subsection{Design Implication on Structuring LLM Explanation Text into Components}

The trust gain resulted from presenting always-on structured explanations may be superficial trust based on simplicity rather than understanding. It’s therefore crucial to build genuine trust and avoid blind reliance on oversimplified icons. Coupled with participants’ desire for sufficient and readable details, accurate summarization and iconification play a critical role in ensuring appropriate understanding of AI recommendations. Designers should carefully balance brevity with clarity when creating icons and summaries for glanceability. As an example, Zender and Mejía \cite{zender2013improving} suggest that adding more symbols can enhance comprehension by providing context and narrow the focus. For instance, adding a ``bookshelf’’ icon to a ``library’’ icon can improve understanding compared to using just a ``man reading’’ icon. They also encourage designers to test icons through multiple iterations and learn from user feedback for improvement. 

There is also likely a continuum of text verbosity beyond simply hiding and displaying explanations that adapts to different tasks, e.g., ``progressively disclosing’’ a continuum of explanation levels \cite{springer2019progressive}. Computational models could be developed to predict the appropriate level of detail based on user preferences, current contexts, and interaction histories. Alternatively, from a design perspective, users should be given control over the amount of explanation detail they receive, ranging from icons and diagrams to unstructured text.

\subsection{Design Implication on Adaptively Presenting Structured Explanations}


Our analysis highlighted that participants preferred AI to be as transparent as possible and found that inefficient adaptive structured explanations were less favored compared to always-on explanations. Thus if adaptive explanations are to be used, it is crucial to carefully select the optimal timing to present and hide explanations. Based on prior literature \cite{xu2023xair}, explanations should be displayed automatically when there is a need to manage AI errors or to resolve user surprise/confusion. More sophisticated methods need to be developed to implement adaptive explanations effectively. 
Model developers should explore more accurate indicators of AI correctness to detect when the AI makes mistakes.
For situations require the understanding of user states, building a personalized user mental model based on interaction logs to predict when users are confused, unfamiliar with the outcome, or uncertain about their own choices \cite{ma2023should, ma2024you, wang2022will}, could further improve the effectiveness of adaptive explanations.

\subsection{Limitations and Future Work} 

This research has several limitations. The results of our study are limited by the specific type of AI assistance scenario (i.e., goal-oriented contextual action recommendation)
and the fact that participants watched third-party videos rather than interacting with an actual working system on a smartwatch. The controlled in-lab setting may have also made participants deliberately curious about seeing more information, \rv{whereas in a longitudinal study conducted in real-world settings, they may spend less time deciding what to do next, and their user preferences may evolve over time}.
Moreover, our LLM recommendation pipeline used only video frames as input. In practice, other signals such as digital states, audio, eye tracking, text recognition, EMG, IMU could all be utilized as input for the modality encoder models and mapped into the LLM embedding space \cite{moon2023anymal} to enable more accurate action inference. 
\rv{Furthermore, the generated LLM recommendations carry potential risks including biases against underrepresented groups and accessibility challenges for populations with disabilities or limited technological proficiency.}

A promising future direction to explore for effectively presenting AI explanations on ultra-small devices is to flexibly leverage multiple modalities and selectively summarize information from various input signals \cite{miller2019explanation}.
With more input signals available such as audio and eye tracking, explanations could incorporate much richer information than the four contextual concepts currently used (i.e., object, activity, goal, location). For example, the AI explanations could include the user's historical preferences, dialogues with friends, or current health state.
In addition to text and icons, there are alternative modalities for displaying explanations that can reduce mental demand while improving the naturalness of information delivery. For example, on interfaces like AR smartglasses, which support overlay displays, real-world objects can be highlighted and overlaid onto the physical environment to enhance understanding \cite{du2020depthlab}.

\section{Conclusion}
In this research, we explored how to effectively present glanceable LLM explanations on ultra-small devices. We proposed that LLM explanations should be \textit{structured} using defined contextual components during prompting to achieve spacial glanceability, and \textit{adaptively} presented based on confidence levels to achieve temporal glanceability. 
We conducted a user study to evaluate the effectiveness of structuring and adaptively presenting explanations on user experiences and perceptions of an AI assistant system using a simulated smartwatch UI. We found that structured explanations reduced users' time to action and lowered perceived mental load when reading the explanations. Always-on structured explanations led to higher acceptance rates of the AI recommendations.  But structured explanations also lacked sufficient and readable details and were thus less satisfying than unstructured explanations. Additionally, while adaptive structured explanations offered participants more control over the interface, they introduced a redundant interaction step and generally hindered effective interaction with the AI, highlighting the need for improvements in the timing for adaptive presentation of AI explanations.
Our study should thus serve as a foundational starting point to understand how users prefer to view AI explanations on ultra-small devices and encourage future research and technical developments that promote responsible AI for users of AI-based personal assistants.


\bibliographystyle{ACM-Reference-Format}
\bibliography{sample-base}

\appendix
\section{LLM Prompts Used}

\subsection{Unstructured Explanation Prompt}
\label{app:prompt_baseline}




{\footnotesize
\begin{spacing}{1.0} 
\texttt{Your input will be (1) a series of narrations of videos taken from a smartglasses user's perspective and (2) objects detected in the user's field of view. These two pieces of information are detected by upstream machine learning models.}

\texttt{Input:}

\texttt{
1) Video narrations, in the order of the least to most recent: 
['\#C C looks around ', '\#C C turns around ', '\#c c walks in the supermarket', '\#C C walks in the shop', '\#C C walks in the supermarket ', '\#C C looks around ', '\#c c walks in the supermarket', '\#C C looks around ', '\#C C walks in the supermarket', '\#C C turns around ', '\#C C turns around ', '\#C C looks around', '\#C C looks around ', '\#C C picks a box from the shelve ', '\#C C operates the phone', '\#C C uses the phone']}

\texttt{2) Detected objects in the user's field of view: 
['bowl', 'bottle', 'oven', 'potted plant', 'orange', 'microwave', 'refrigerator', 'cup', 'book', 'sink', 'toothbrush', 'cell phone', 'broccoli', 'person', 'scissors', 'cake', 'dining table', 'knife', 'couch', 'apple', 'handbag']}

\texttt{The recommendation is "open a pantry organization tutorial on the Youtube app". 
Please provide an <explanation> for this recommendation within 25 words. In the explanation, only focus on referencing information from the contextual input and please think step by step. Please do not restate the recommendation in your explanation.
}
\end{spacing}
}







\end{document}